\documentclass[a4paper]{jpconf}
\usepackage{graphicx}
\begin{document}
\title{Simulation of Cosmic Ray $\nu$ Interactions in Water}

\author{T. Sloan}

\address{ University of Lancaster.\\
Talk presented on behalf of the ACoRNE Collaboration at the ARENA Workshop 2006}

\ead{t.sloan@lancaster.ac.uk}

\begin{abstract}
The program CORSIKA, usually used to simulate extensive cosmic ray air 
showers, has been adapted to a water medium in order to study the acoustic 
detection of ultra high energy neutrinos. Showers in water from incident 
protons and from neutrinos have been generated and their properties are 
described. The results obtained from CORSIKA are compared to those from 
other available simulation programs such as Geant4.
\end{abstract}

\section{Adaptation of the CORSIKA program to a water medium}
The air shower program, CORSIKA (version 6204) \cite{CORSIKA}, has 
been adapted to run 
in sea water i.e. a medium of constant density of 1.025 g per cm$^3$ 
rather than the variable density needed for an air atmosphere. Sea 
water was assumed to consist of a medium in which $66.2\%$ of the 
atoms are hydrogen, $33.1\%$ of the atoms are oxygen and $0.7\%$ of the 
atoms are made of common salt, NaCl. The salt was assumed to be 
a material with atomic weight and atomic number A=29.2 and Z=14, the mean 
of sodium and chlorine. The purpose of this is to maintain the structure 
of the program as closely as possible to the air shower version 
which had two principal atmospheric components (oxygen and nitrogen) 
with a trace of argon. The presence of the salt component had an
almost undetectable effect on the behaviour of the showers. 

Other changes made to the program to accomodate the water medium 
include a modification of the stopping power formula to allow for 
the density effect in water. This only affects the energy loss in 
hadrons since the stopping powers for electrons are part of the 
EGS \cite{EGS} package which is used by CORSIKA to simulate the 
propagation of the electromagnetic component of the shower. 
Smaller radial binning of the shower was also required since shower radii 
in water are much smaller than those in air. In addition the threshold 
for the LPM effect \cite{LPM}, which suppresses pair production from photons 
and bremsstrahlung from electrons at high energy, was reduced to 
the much lower value necessary for water. Similarly, the 
interactions of $\pi^0$s had to be simulated at lower energy than in air 
because of the higher density water medium.  
In all about 100 detailed changes needed to be made to the CORSIKA 
program to accomodate the water medium.

\begin{figure}[h]
\includegraphics[width=36pc]{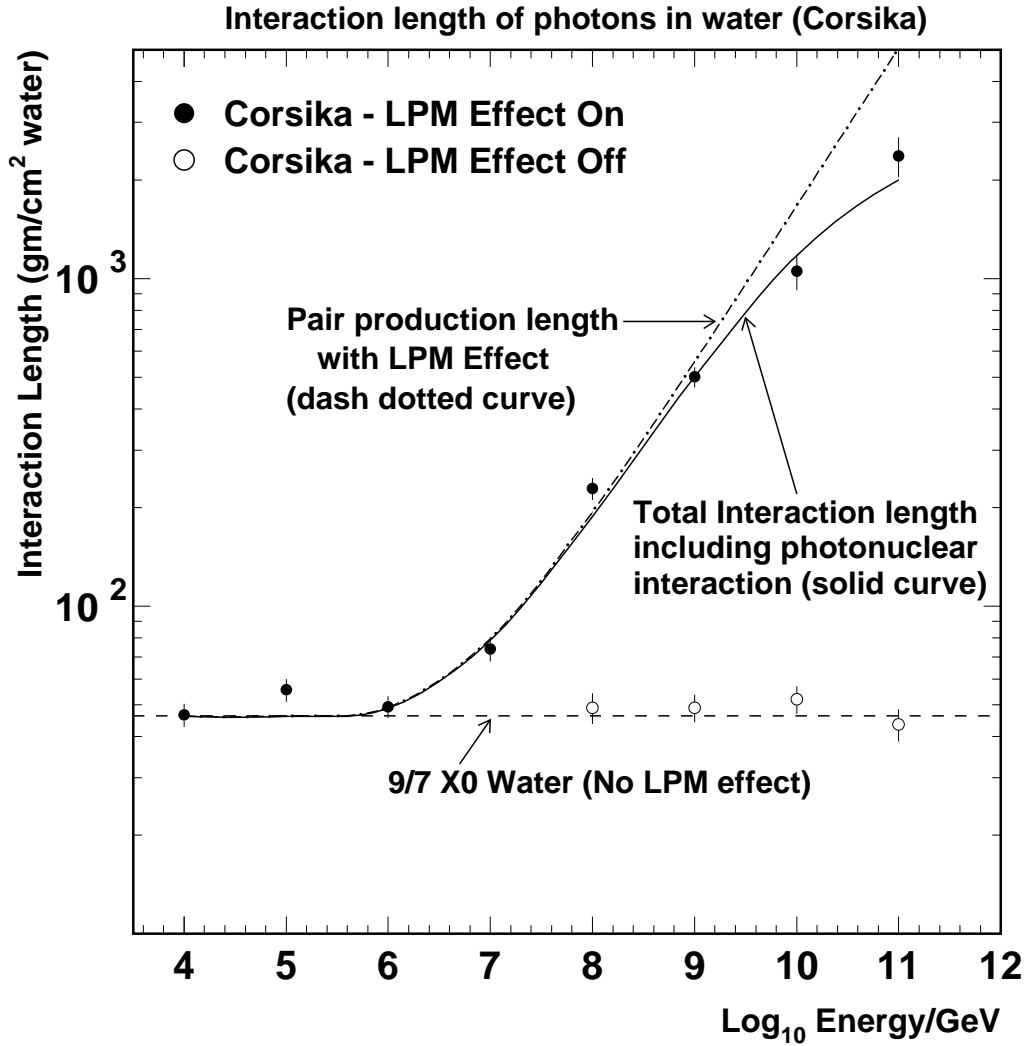}\hspace{2pc}%
\caption{\label{LPMfig}The interaction length 
for high energy gamma 
rays versus the photon energy measured in CORSIKA (data points with 
statistical errors). The dash dotted curve shows the pair production 
length computed from the LPM effect using the formulae of 
Migdal \cite{LPM}. The solid curve shows the computed total interaction 
length, including both pair production and photonuclear interactions 
computed using the values of the photonuclear cross 
section from CORSIKA.}
\end{figure}

To test the implementation of the LPM effect \cite{LPM} in the program 
100 showers from incident gamma ray photons were generated and the mean 
depth of the first interaction (the mean free path) calculated.  
The observed mean free path was found to be in agreement with the 
expected behaviour when both the suppression of pair production 
and photonuclear interactions were taken into account 
(see figure \ref{LPMfig}). This showed that 
the LPM effect had been properly implemented in CORSIKA.  

Considerable fluctuations between showers occurred giving the following 
observed values of the ratios of the root mean square deviations 
to the mean value in proton showers: rms peak energy deposit to  
the peak energy deposit was observed to be $14\%$ at $10^5$ GeV reducing 
to $4\%$ $10^{11}$ GeV, that for the depth of the peak position varied   
from $19\%$ to $7.4\%$ and for the shower width from $63\%$ to $18\%$.
To smooth out such fluctuations averages 
of 100 generated showers will be taken in the following. The statistical 
error on the averages is then given by these RMS values divided by 
$\sqrt 100$. The hadronic energy contributes only about 10$\%$ to the 
shower energy at the shower peak, the remainder being carried by the 
electromagnetic part of the shower. This is a well known effect in 
calorimeters.

\section{Comparison with Other Simulations}

\subsection{Comparison with Geant4}

Proton showers were generated in sea water using the program 
Geant4 (version 8) \cite{Geant} and compared with those generated in CORSIKA. 
Unfortunately, the range of validity of Geant4 physics models for hadronic 
interactions does not 
extend beyond an energy of $10^5$ GeV. Hence the comparison is restricted 
to energies below this.  

\begin{figure}[h]
\includegraphics[width=36pc]{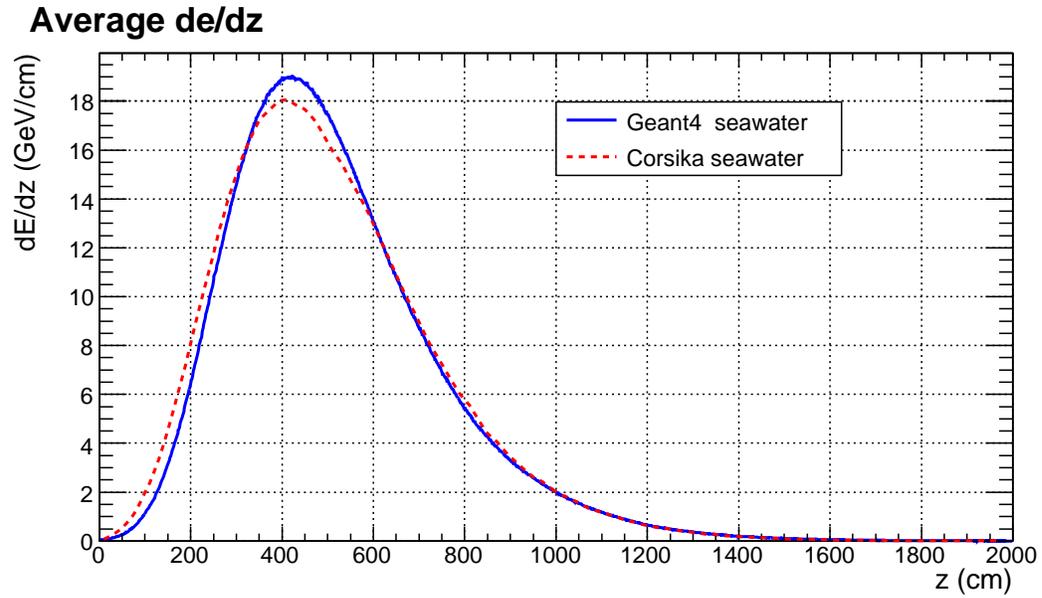}\hspace{2pc}%
\includegraphics[width=36pc]{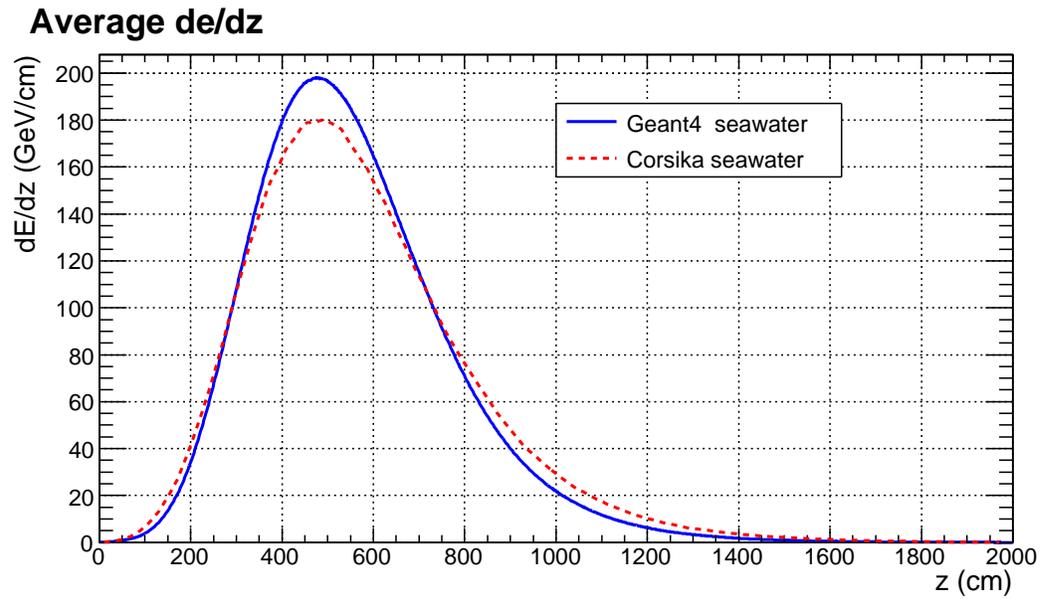}\hspace{2pc}%
\caption{\label{long}Averaged longitudinal energy deposited per unit 
path length of 100 proton showers at energy $10^4$ GeV (upper plot) 
and $10^5$ GeV (lower plot) generated in 
Geant4 and Corsika versus depth in the water. }
\end{figure}

\begin{figure}
\begin{minipage}{20pc}
\includegraphics[height=50pc,width=21pc]{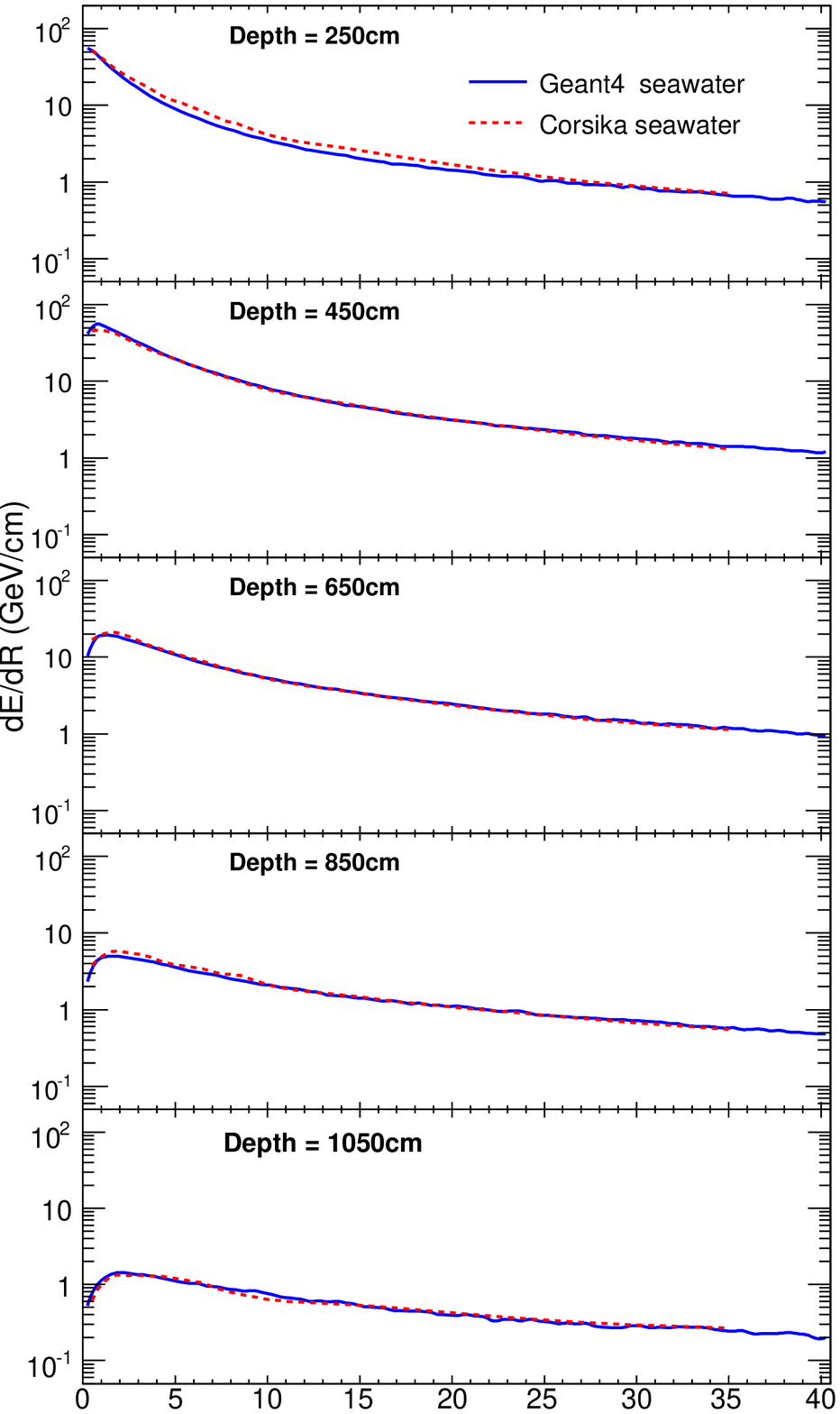}
\end{minipage}\hspace{-1pc}%
\begin{minipage}{20pc}
\includegraphics[height=50pc,width=21pc]{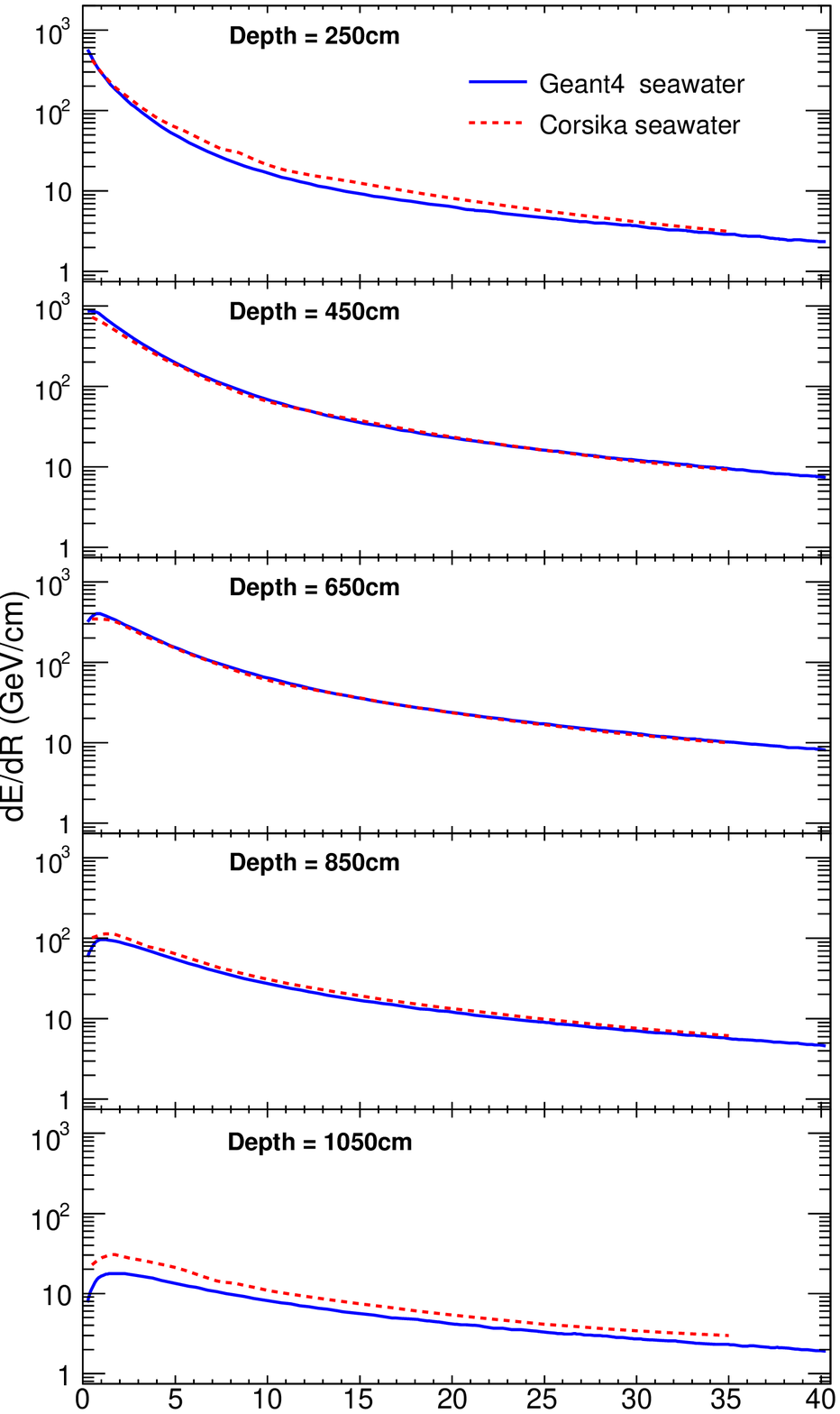}
\end{minipage}\hspace{-1pc}
\vspace*{-5mm}
\caption{\label{rad}Averaged radial energy deposited per 20 gm cm$^{-2}$ 
vertical slice per unit radial distance for 100 proton showers at 
energy $10^4$ GeV (left hand plot) and $10^5$ GeV (right hand plot) 
generated in Geant4 and CORSIKA versus distance from 
the axis in the water for different depths of the shower.}

\end{figure}

Figs. \ref{long} show the longitudinal distributions of proton showers 
at energies of $10^4$ and $10^5$ GeV (averaged over 100 showers) 
as determined from Geant4 and CORSIKA. The showers from CORSIKA tend to 
be slightly broader and with a smaller peak energy than those generated 
by Geant4. The difference in the peak height is $\sim 5\%$ at $10^4$ GeV 
rising to $\sim 10\%$ at energy $10^5$ GeV. Figs. \ref{rad} show the 
radial distributions. The differences in the longitudinal distributions 
are reflected in the radial distributions. However, the shapes of the 
radial distributions are very similar between Geant4 and CORSIKA. 
CORSIKA produces somewhat less ($\sim 20\%$) energy than Geant4 near 
the shower axis at depths between 450 and 850 g cm$^{-2}$ where most 
of the energy is deposited. The acoustic signal from a shower is most 
sensitive to the distribution near the axis ($r \sim 0$).   

\subsection{Comparison with Alvarez-Muniz and Zas Simulation}

The CORSIKA simulation was also compared with the longitudinal shower 
profile computed in the simulation by Alvarez-Muniz and Zas \cite{AZpaper}.
There was a reasonable agreement between the longitudinal shower shapes 
from CORSIKA and 
those shown in fig. 2 of ref.~\cite{AZpaper}. However, the number of electrons 
at the peak of the CORSIKA showers was $\sim 20\%$ lower than those from 
ref.~\cite{AZpaper}. 
Their procedure involves a fast hybrid Monte Carlo which simulates one 
dimensional showers down to a cross over energy below which  
parameterisations are used. 
The total number of electrons produced is sensitive 
to the lower energy down to which the simulation proceeds, which is not 
specified in the paper. Given this unknown, the agreement between 
CORSIKA and the simulation of ref.~\cite{AZpaper} is probably satisfactory.     

\section{Comparison of Radial Distributions and Published Parameterisations}
\begin{figure}
\begin{minipage}{20pc}
\includegraphics[height=50pc,width=21pc]{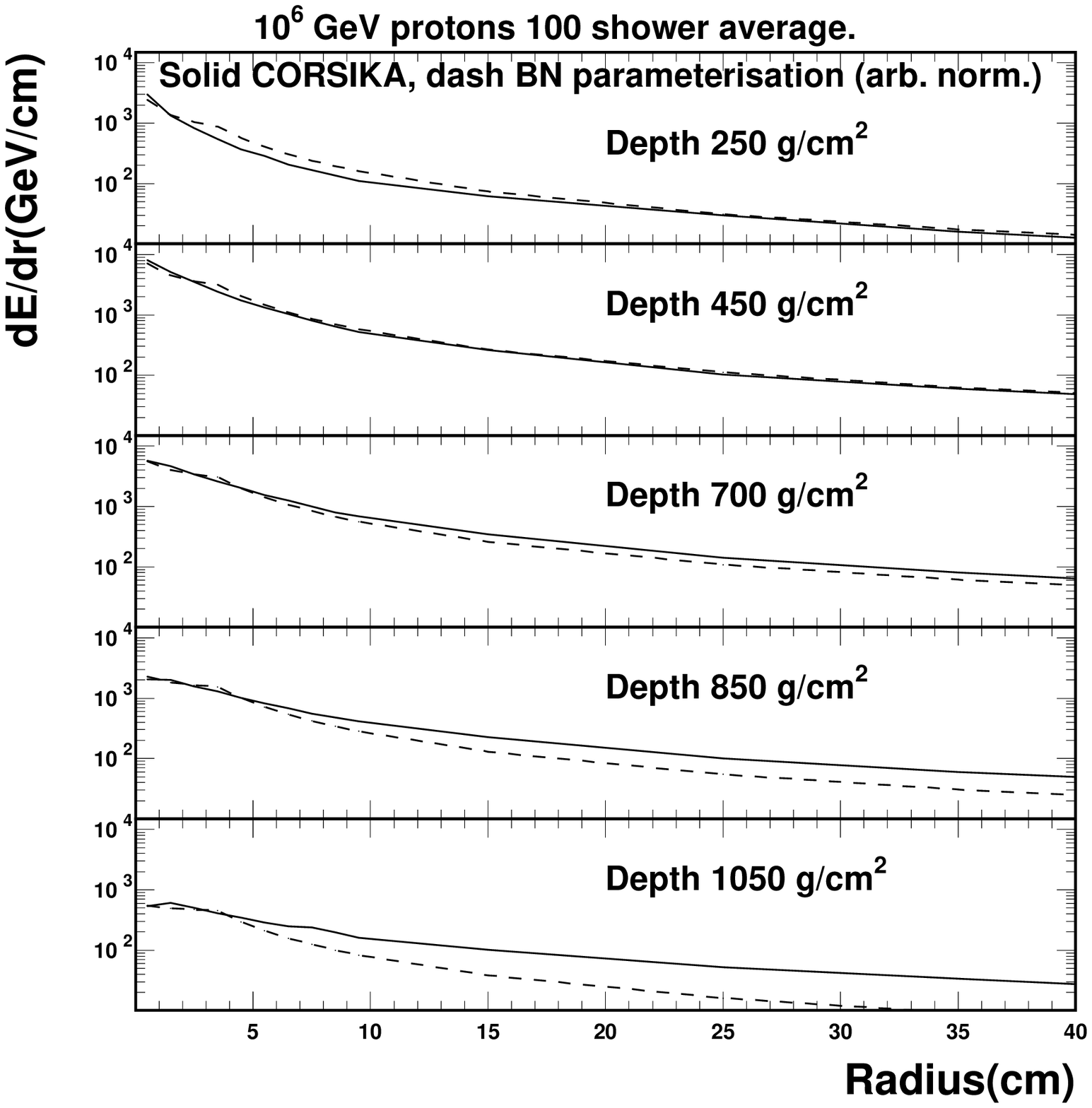}
\end{minipage}\hspace{-1pc}%
\begin{minipage}{20pc}
\includegraphics[height=50pc,width=21pc]{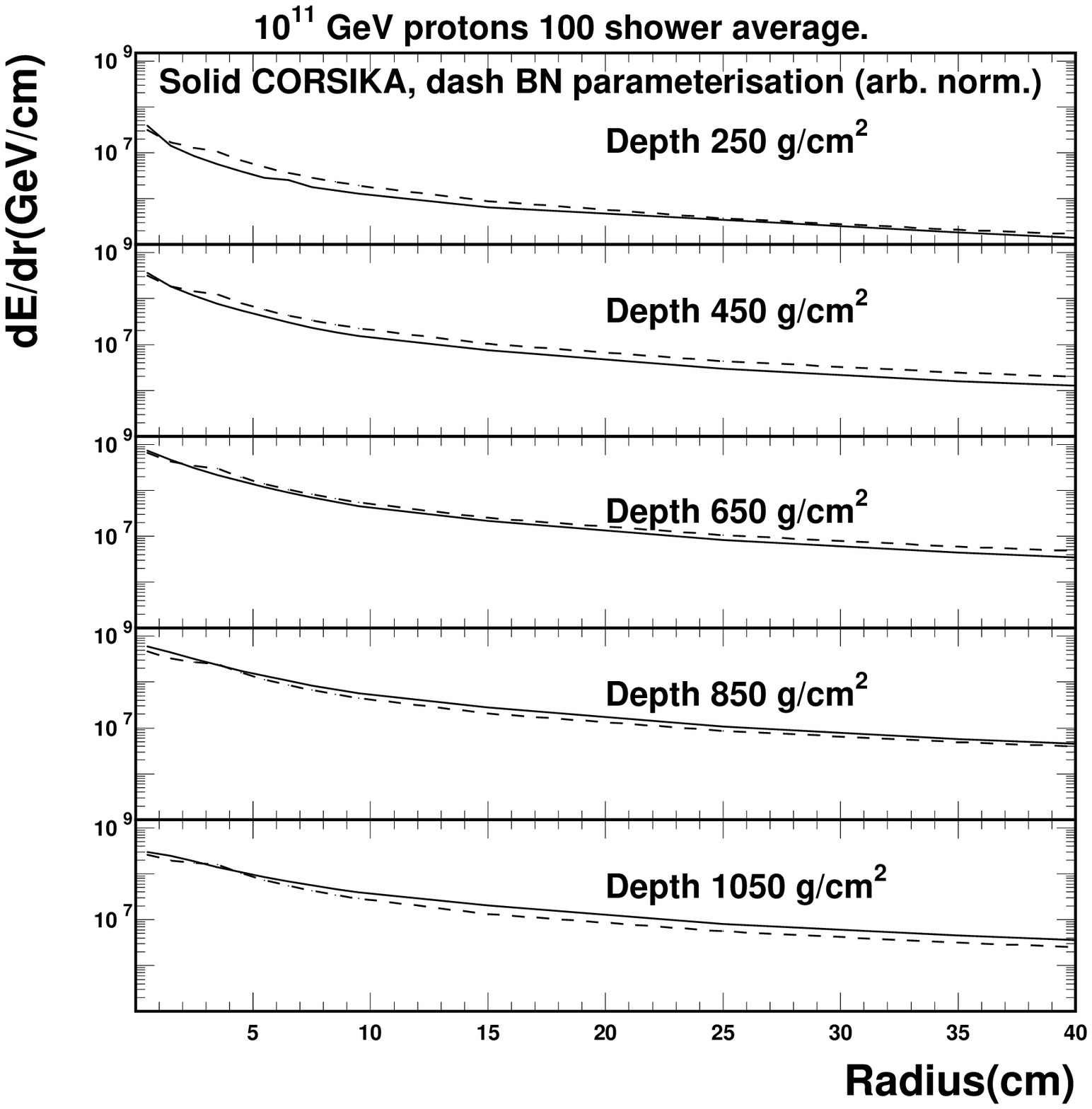}
\end{minipage}\hspace{-1pc}
\vspace*{-5mm}
\caption{\label{bn}Averaged radial energy deposited per 20 gm cm$^{-2}$ 
vertical slice per unit radial distance for 100 proton showers at 
energy $10^6$ GeV (left hand plot) and $10^{11}$ GeV (right hand plot) 
in water compared to the Niess-Bertin parameterisation \cite{Valentin}.}

\end{figure}

Niess and Bertin \cite{BN,Valentin} parameterise the radial density 
distribution of showers in water as 
\begin{equation}
\rho(r)=k x^n 
\end{equation}
where $k$ is a normalising constant chosen arbitrarily here at each depth, 
$x=3.5/r$ and $n=1.66-0.29 z/z_{max}$ for $r < 3.5$ cm or 
$n=2.7$ for $r > 3.5$ cm. This is the parameterisation for pion showers 
given in \cite{Valentin}. Here $z$ is the depth and $z_{max}$ 
is the depth of maximum energy deposition. The energy deposited per cm 
at radius $r$ is then proportional to $r\rho(r)$. Fig \ref{bn} shows the 
radial distribution from CORSIKA compared with this parameterisation.
There is reasonable agreement between them particularly at small 
values of $r$. Hence the two should predict similar distributions of 
the frequencies of acoustic signals from the hadron shower in water.   

\begin{figure}
\begin{minipage}{20pc}
\includegraphics[height=50pc,width=21pc]{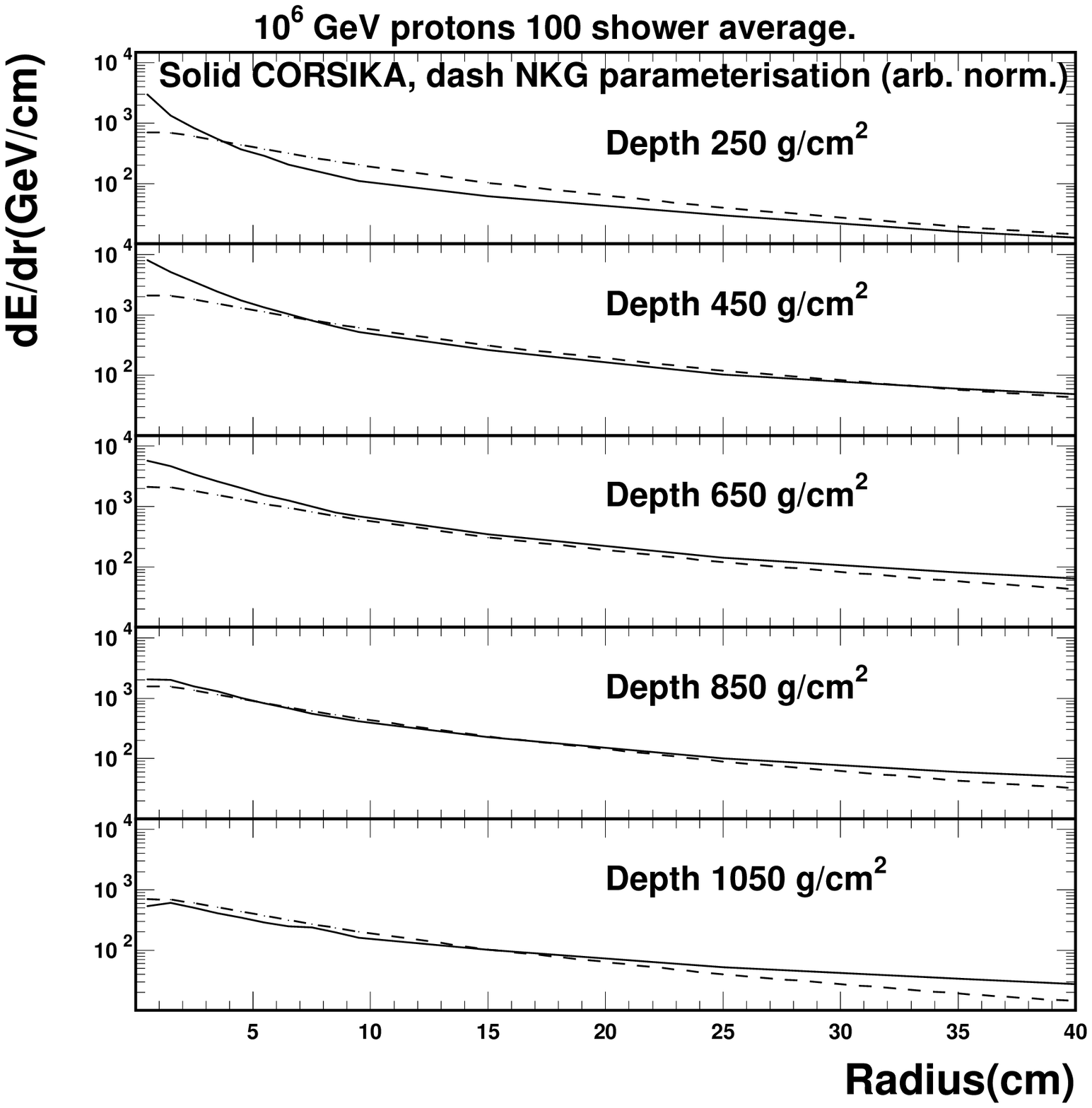}
\end{minipage}\hspace{-1pc}%
\begin{minipage}{20pc}
\includegraphics[height=50pc,width=21pc]{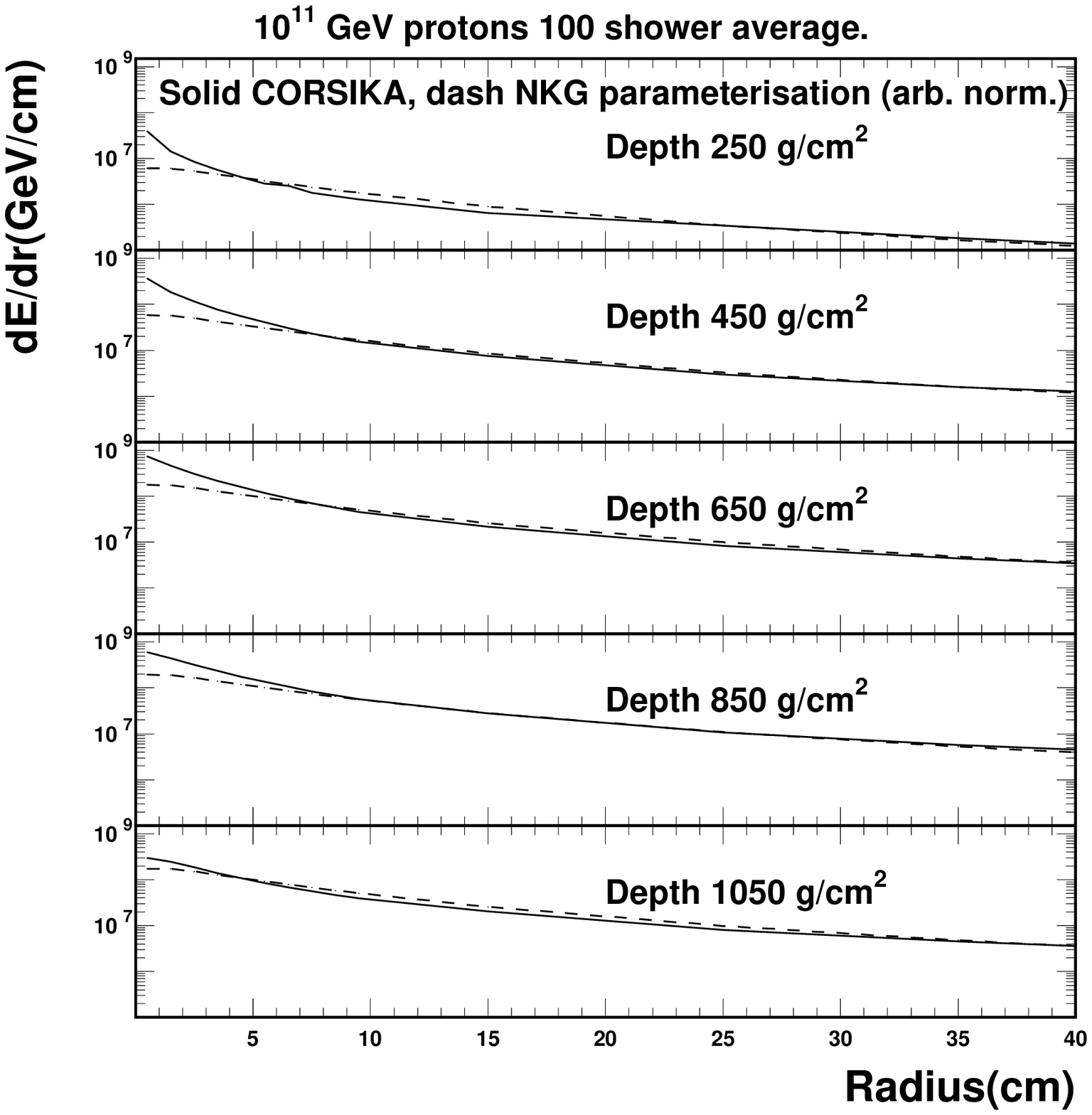}
\end{minipage}\hspace{-1pc}%
\vspace*{-5mm}
\caption{\label{nkg}Averaged radial energy deposited per 20 gm cm$^{-2}$ 
vertical slice per unit radial distance for 100 proton showers at 
at energy $10^6$ GeV (left hand plot) and $10^{11}$ GeV 
(right hand plot) in water compared to the parameterisation used 
by the SAUND Collaboration \cite{Justin}.}
\end{figure}

The SAUND Collaboration \cite{SAUND} uses the following parameterisation 
\cite{Justin} for the energy deposited per unit depth, $z$, and per 
unit annular thickness at radius $r$ from a shower of energy $E$
\begin{equation}
\frac{d^2E}{drdz}=E k (\frac{z}{z_{max}})^t \exp{(t-z/\lambda)}~ 2 \pi r \rho(r)
\end{equation}
where $z_{max}=0.9 X_0 \ln(E/E_c)$ is the maximum shower depth,  
$X_0 =36.1$ g cm$^{-2}$ is the radiation length and $E_c = 0.0838$ 
GeV is the critical energy in water. The constants 
$t=z_{max}/\lambda$ where $\lambda = 130 - 5 \log_{10}(E/10^4 \mathrm{GeV})$ 
g cm$^{-2}$ and $k=t^{t-1}/\exp{(t)}\lambda \Gamma(t)$. 
The radial density is given by  
\begin{equation}
\rho(r)=\frac{1}{r_M^2}a^{s-2} (1+a)^{s-4.5} \frac{\Gamma(4.5-s)}{2\pi \Gamma(s) \Gamma(4.5-s)}
\end{equation}
where $a=r/r_M$ with $r_M=9.04 $  g cm$^{-2}$ the Moliere radius in water 
and $s=1.25$.  
Fig \ref{nkg} shows the 
radial distributions from CORSIKA compared with the absolute predictions of 
this parameterisation. There is qualitative agreement between them.  
However, CORSIKA predicts relatively more energy at small 
$r$ i.e. a harder frequency spectrum for acoustic signals.   

\section{Simulation of $\nu$ induced Showers}

The CORSIKA program has an option to simulate the interactions 
of neutrinos at a fixed point \cite{Ofelia}. The first interaction 
is generated by the HERWIG package \cite{Herwig}. Some problems 
have been encountered in simulating $\nu$ showers at energies 
above $4~10^7$ GeV. However, the package seems to work well at 
energies below this. At these energies the mean value of the energy 
transfered to the hadrons in the interaction is about $25\%$ of the 
incident neutrino energy \cite{Ghandi} and this fluctuates in 
different interactions between zero and $100\%$ 
with an RMS value of $25\%$. Hence much larger fluctuations 
occurred for neutrino than for proton showers. 
The shapes of the hadron shower profiles for neutrinos were observed to 
be similar to those generated by proton showers in water at an energy 
of $25\%$ of the neutrino energy.     

\section{Conclusions}
The program CORSIKA has been modified to work in a water medium. Hadron and 
electromagnetic showers can be generated routinely. Neutrino interactions 
at low energy can be generated by the HERWIG package interfaced to 
CORSIKA  and it may be possible to extend this to higher energies. 

\subsection{Acknowledgments}
I wish to thank all my colleagues in the ACoRNE Collaboration for their 
help and support. Special thanks go to Jon Perkin who produced several 
of the plots shown in this talk. I should also like to thank Ralph Engel, 
Dieter Heck, Johannes Knapp and Tanguy Pierog for their assistance in 
modifying the CORSIKA program.

\vspace*{-1mm}



\vspace*{-1mm}

\vspace*{-1mm}

\end{document}